# Experimental study of cavitation development in liquid in pulsed non-uniform electric field under the action of ponderomotive forces

Andrey Yu. Starikovskiy and Mikhail N. Shneider

*Department of Mechanical and Aerospace Engineering, Princeton University, Princeton, NJ, 08544*

**Abstract**

In this paper, the Rayleigh scattering method is used to study the formation of cavitation in water in a pulsed inhomogeneous electric field when a nanosecond high voltage pulse is applied to a needle-like electrode. The observational results confirm the theoretical picture of cavitation development under the action of electrostrictive ponderomotive forces. The values of the negative pressure, the average size of the cavitation nanovoids and their concentration were obtained from these measurements, which are in agreement with theoretical estimates.

**Introduction**

The origin of cavitation and the dynamics of its development is one of the fundamental problems of modern hydrodynamics. Understanding the conditions for the initiation of cavitation and the dynamics of its development is necessary to solve many problems in science and technology. For example, knowledge of the critical parameters of cavitation initiation and its characteristics is important for solving the problem of propeller erosion in shipbuilding. However, there is a very large scatter of data in the literature on the conditions and critical characteristics of the onset of cavitation [1]. This is due to the uncertainty in the values of negative pressure and its duration under different experimental conditions. In addition, direct observation of the onset of cavitation is extremely difficult due to the small size of the cavitation discontinuities at the initial stage of their evolution (R ~ 15 nm).

Experiments [2] showed that when a nanosecond pulse voltage is applied to a needle-like electrode in water, discharge occurs at electric fields that are much lower than expected from estimates of the electric strength of water based on its density. The puzzle was that the breakdown in the liquid occurred without the formation of gas bubbles in which electrons could gain enough energy to ionize water molecules. Bubbles would not have been able to form in a time on the order of 1-5 ns, and they were not observed in the experiment. A bubble-free breakdown mechanism in liquid dielectrics was proposed in [3] and further developed and experimentally studied by many authors [4-13]. However, to the best of our knowledge, direct measurements of the cavitation initiation process and its parameters have not yet been performed.

The values of negative pressure in a liquid are much better determined in non-uniform electric fields with known electrode geometry and voltage pulse. In this case, tensile electrostrictive (ponderomotive) stresses leading to the development of cavitation in a liquid dielectric are fully known. At the same time, the appearance of cavitation voids can be observed by changes in the scattering of laser radiation passing through the liquid medium where cavitation is developing. Since, for the wavelength of the laser radiation $\lambda$, $R/\lambda \ll 1$ is satisfied, it has been previously proposed to observe the Rayleigh scattering on the formed cavitation nanovoids, from which it is possible to estimate the characteristic sizes of the nanovoids and their density [15,16]. Moreover, for these experimental conditions, it is possible to determine with good accuracy the critical values of the negative pressure, determined by the voltage on the tip electrode and its geometry. The possibility of such observation of incipient cavitation and diagnostics of its parameters has been demonstrated for the first time in this study.

It is known that the development of cavitation nanovoids in the region of electrostrictive negative pressure is a necessary condition for the development of nanosecond electrical breakdown in water [16]. However, in this work we will not consider issues related to ionization and the development of an electric discharge, but will focus only on cavitation and the measurement of its characteristic parameters.

**Experiment**

The study of the development of the discharges in the liquid was carried out using a pulse voltage generator FID30-20. The generator created voltage pulses with a rise time of 600 ps, an amplitude of 30 kV, and a half-height duration of 23 ns on the point-to-plane discharge gap, completely immersed in distilled water. The pulse voltage was monitored by a back-current shunt embedded in the braid of a high voltage coaxial cable. The interelectrode distance was 5 mm. The high voltage needle electrode was made of tungsten. The electrode was conical and angled at 14 degrees. The radius of curvature of the electrode tip was 0.5 μm, and the diameter of the cylindrical part was 0.52 mm. The low voltage flat electrode was made of copper. The discharge chamber windows were made of 1.5 mm thick fused quartz.

A BNC-577-AT45 delayed pulse generator provided the sync signal for the high voltage generator and generated a 7 ns current pulse for the L520P50 laser diode. The laser diode produced a pulse of radiation at a wavelength of 520 nm, which was focused by a short focal length lens onto a 50 μm diameter aperture, then a 100 mm focal length lens produced a parallel beam of light which was used to obtain a shadowgraph of the discharge development region. The laser diode radiation passing through the discharge region was focused by a quartz lens at 5x magnification onto the photocathode of the image intensifier of the PicoStar HR12 ICCD camera, which operated with a 2 ns gate in the case of shadowgraph imaging, and with a 500 ps gate in the case of the discharge imaging. The camera photocathode had an S20 spectral response with sensitivity from 200 to 650 nm. Synchronization of the CCD camera and the image intensifier with the discharge and laser diode emission pulse was provided by a second delayed pulse generator BNC-577. The synchronization accuracy of all three systems (the high voltage generator, the laser radiation source and the ICCD camera) was better than 400 ps and was limited by the start-up jitter of the different units. The schematic of the setup and its general view are shown in Figure 1.

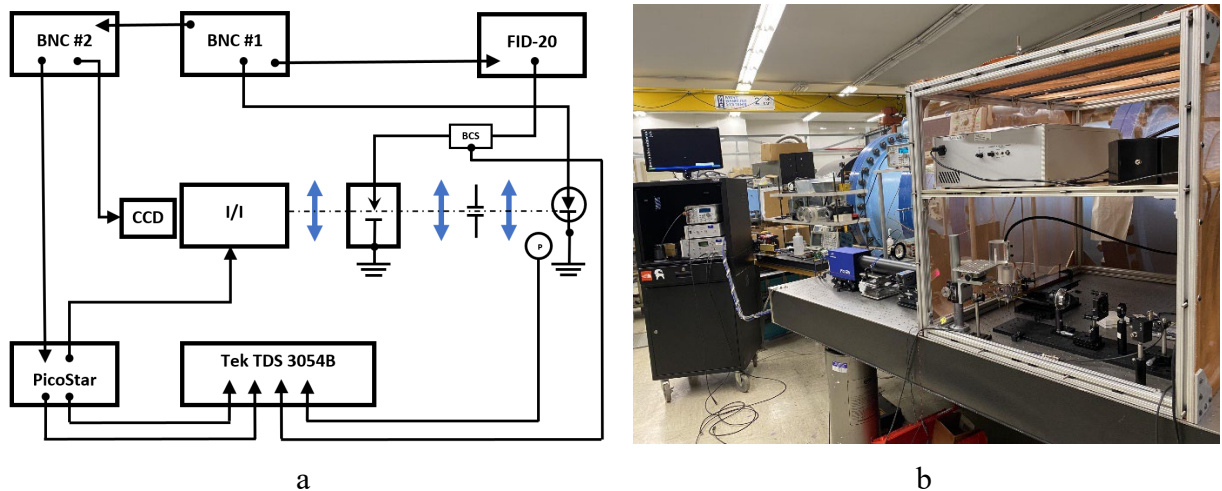

Figure 1. Schematic (a) and general view of the installation (b).

Thus, the setup allowed us to study the evolution of the nanosecond discharge in the liquid with a time resolution better than 500 ps during the breakdown phase by discharge radiation and better than 2 ns in the afterglow phase by shadowgraph imaging.

**Discharge development**

The main part of the discharge radiation in water falls on the lines of atomic hydrogen and OH radicals excited by electron impact in the region of high electric field. The strongest electric field at the initial moment is formed near the tip of the high voltage electrode (Figure 2).

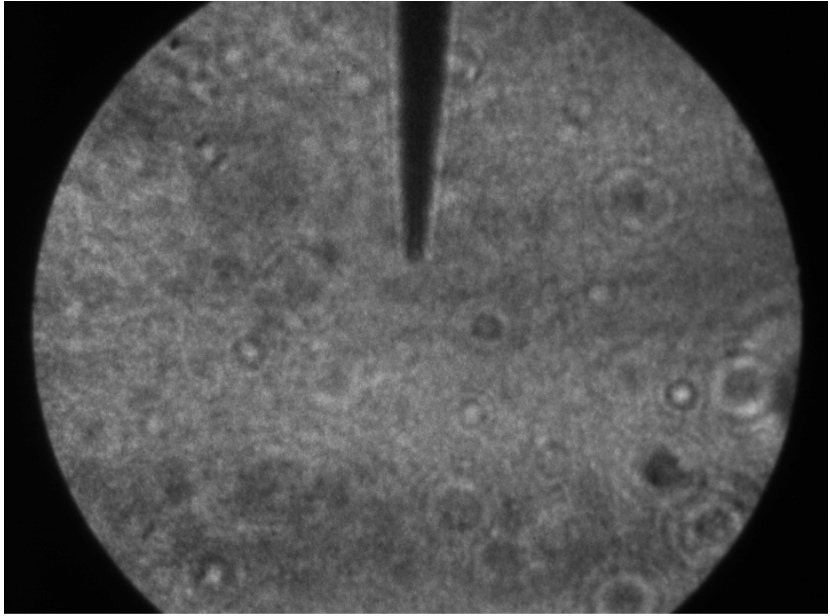

Figure 2. Shadowgraph image of the tip of a high voltage electrode immersed in water. The image size is 9.0×6.7 mm$^2$. Camera gate is 2 ns, illumination by a laser diode at a wavelength of 520 nm. Averaging over 120 frames.

Figure 3 shows the sequence of images of the development of a high voltage nanosecond discharge near the electrode according to the emission detected by the ICCD camera with an exposure time of 500 ps. It can be seen that the discharge starts almost instantaneously due to the fast voltage rise in the pulse and the high collision frequency in the dense medium. The size of the emitting region and the emission intensity increase until the 8$^{th}$ nanosecond and then remain at a constant level for the next 5 nanoseconds. At the interval 13-17 ns, the pause of discharge luminescence associated with the expansion of the ionized region, the accumulation of space charge and the decrease of electric fields near the high voltage electrode is clearly visible. The 18-24 ns interval corresponds to the trailing edge of the high voltage pulse. The electric field in this interval changes direction due to the decrease in potential of the high voltage electrode, and a reverse breakdown wave occurs, collecting the space charge back into the high voltage electrode.

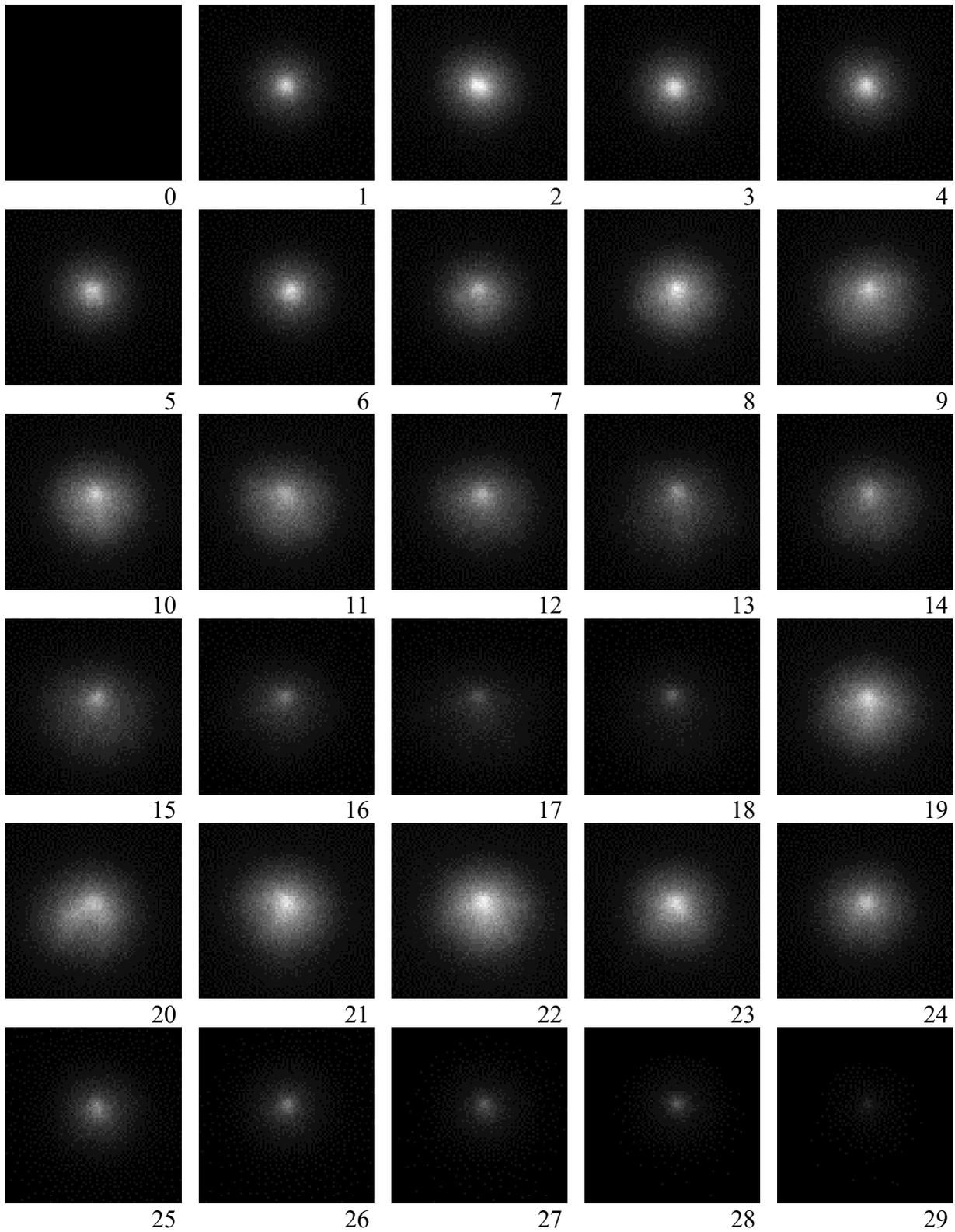

Figure 3. Dynamics of nanosecond discharge evolution in water near the tip of the high voltage electrode. The image size is 2.73×2.73 mm$^2$. Camera gate is 500 ps, single exposure. Pulse voltage is 30 kV. Numbers below images indicate time in nanoseconds.

More importantly, the spatial distribution of the radiation variation for a single discharge is practically homogeneous, indicating the absence of a pronounced streamer structure of the discharge region. Such a high degree of discharge homogeneity results from the high rate of the voltage increase at the high voltage electrode (30 kV for 600 ps) and the small radius of the electrode tip (0.5 μm), which allowed the ionization wave to start simultaneously from the entire electrode surface in all directions.

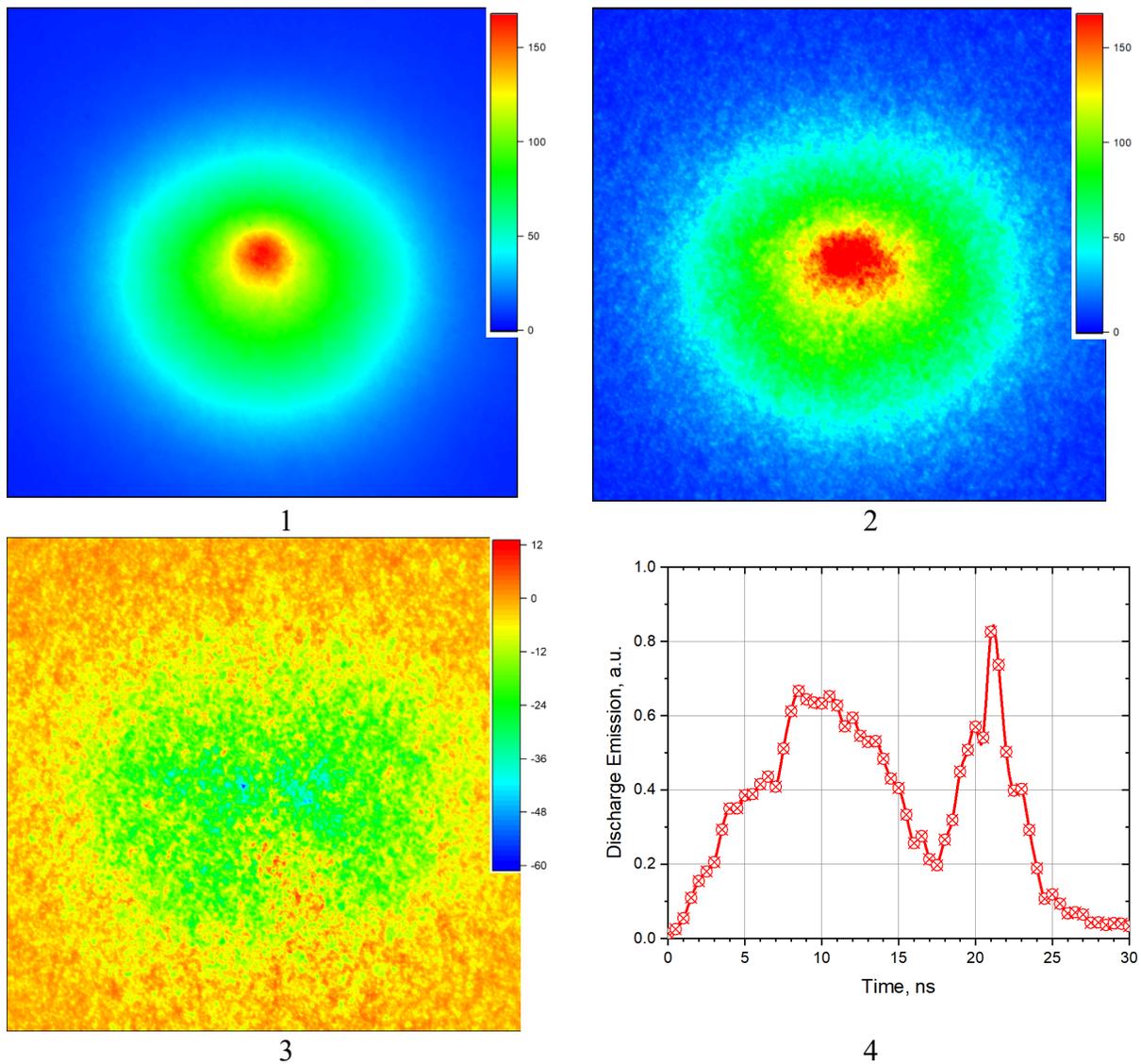

Figure 4. 1-3: Structure of the central region of the discharge corresponding to the time t = 10 ns after the start of the pulse. 1 - Radiation distribution averaged over 110 frames. 2 - Instantaneous radiation distribution. 3 - Difference between averaged and instantaneous distribution. Camera gate is 500 ps, image size 2.2×1.7 mm$^2$. 4 - Radiation dynamics of the whole discharge area as a function of time.

The integral dynamics of the discharge radiation reflects the picture of its development described above (Fig. 3). The pronounced minimum of the discharge emission at the 18th nanosecond is explained by the accumulation of spatial charge around the electrode. The subsequent reverse breakdown wave causes a second luminescence peak at the 21st nanosecond (Figure 4.4).

**Formation of nanovoids in the zone of discharge development**

High electric fields near the tip of the high voltage electrode cause significant ponderomotive forces in this region. The negative pressure generated in this region is equal to [16]:

$$|P_E| = 0.5\varepsilon_0 E^2 \frac{\partial \varepsilon}{\partial \rho}\rho \approx 0.5\alpha\varepsilon_0\varepsilon E^2 \qquad (1)$$

where $E$ is the electric field, $\varepsilon_0$ is the dielectric permittivity of vacuum, $\varepsilon$ is the dielectric permittivity of water $\varepsilon = 81$. In the case of water, the parameter $\alpha \approx 1.5$ [16]. The magnitude and distribution of the electric field near the electrode depend on the size and geometry of the electrode. In the simplest approximation, if the electrode is considered as a long conducting cylinder with a hemispherical tip of radius $r_{el}$, the electric field near the tip is given by the simple formula [17]:

$$E = Ur_{el}/(2r^2). \qquad (2)$$

Near the tip of the electrode, a reasonably good approximation would be, $E \sim \frac{U}{2r}$. Apparently, the same approximation can be used near the head of the streamer. For a typical streamer channel radius of about $r = 50$ μm ($d \sim 100$ μm) under our conditions (see discussion below) this leads to an estimate of the value of the pressure generated by the ponderomotive forces at a potential of the high voltage electrode of 30 kV ($E \sim 3 \times 10^8$ V/m), $|P_E| \sim 48.4$ MPa. The compressibility of water in this pressure range is nearly constant at $\beta = -\frac{1}{V}\left(\frac{\partial V}{\partial P}\right) = 0.47 \times 10^{-9}$ Pa$^{-1}$. Taking this into account, we can expect a relative decrease of the liquid volume in the region of high electric fields up to

$$\frac{\Delta V}{V_0} \sim \beta|P_E| \sim 2 \div 3\%$$

depending on the magnitude of the applied voltage pulse (Figure 5).

Such a decrease in the volume occupied by the liquid phase in nanosecond timescales cannot be compensated by the flow of liquid from other regions, and leads to the formation of voids. The typical size of such voids is tens of nanometers, making it possible to diagnose their presence by Rayleigh scattering of radiation from an external source.

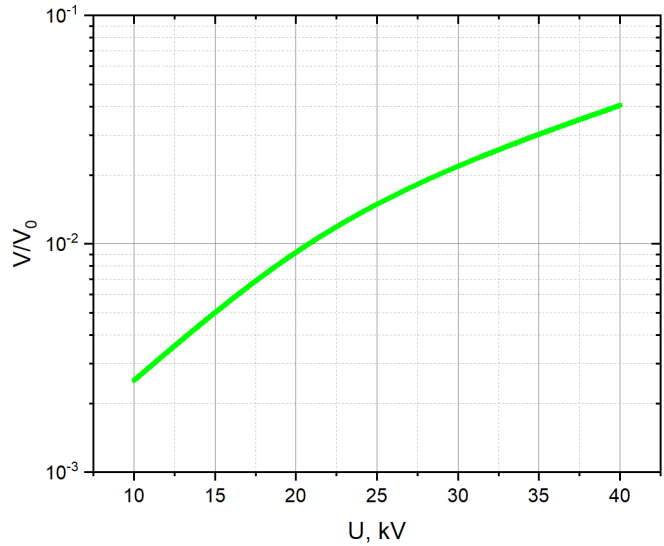

Figure 5. Relative decrease of liquid volume in the region of large electric fields under the influence of ponderomotive forces. The streamer channel radius is assumed to be $r = 50$ μm.

In Figure 6.1, the emission scattering region of the laser diode radiation (520 nm) in the discharge development zone is clearly visible (compare with Figure 2). The size of the radiation scattering region correlates well with the discharge radiation region at the time of maximum streamer propagation ($t = 21$ ns after discharge start, Figure 6.2).

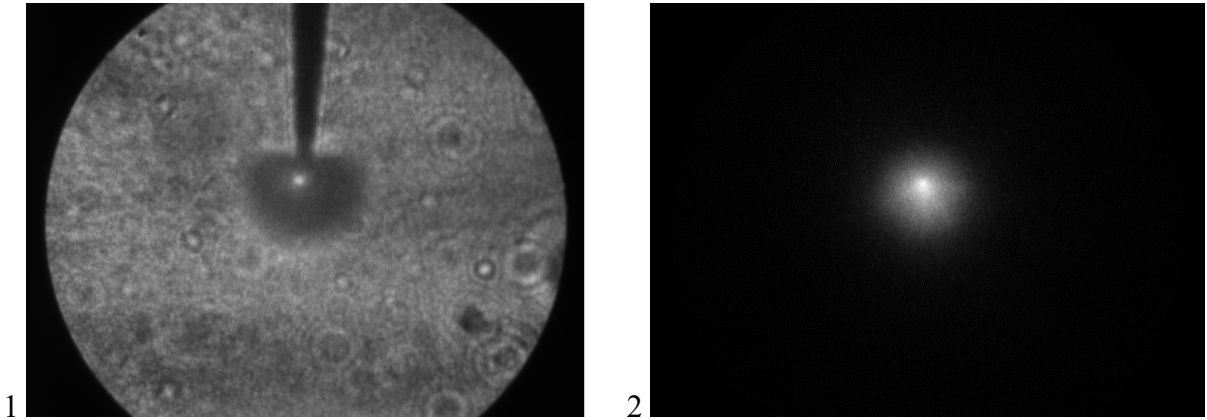

Figure 6. 1 - Shadowgraph image of the discharge development zone 60 ns after the start of the high voltage pulse. 2 – Discharge emission at the moment of maximum streamer propagation ($t = 21$ ns). The image size is 9.0×6.7 mm². Camera gate is 2 ns, illumination by a laser diode at 520 nm wavelength. Averaging over 130 frames.

The distribution of radiation scattering averaged over many high voltage pulses does not allow us to analyze the details of the discharge structure. Therefore, Figure 7 shows instantaneous shadowgraphs of the radiation scattering near the high voltage electrode obtained with a camera gate of 2 ns without averaging over different events. In the shadowgraphs corresponding to the time instants of 45 and 50 ns, the residual radiation of the discharge near the tip of the high voltage electrode is clearly visible. Later, this radiation practically disappears and the whole region occupied by the streamer channels becomes dark due to Rayleigh scattering of radiation on nanovoids formed by ponderomotive forces. In contrast to Figure 6.1, the structure of the discharge is clearly visible in the images without event averaging. In particular, the typical radius of the heads of individual ionization waves, clearly visible in the shadowgraph images of Figure 7, is approximately $r \sim 50$ μm. This estimate allows us to calculate the magnitude of the pressure generated by the ponderomotive forces (equation (1)).

The reduction of the volume of the liquid phase under the action of ponderomotive forces and the formation of nanovoids are related, since the total volume of the cavities is equal to the change in the volume of the liquid. A rough estimate of the relationship between the size of nanovoids R and their concentration in the zone of action of ponderomotive forces n can be obtained by approximating the quasi-spherical shape of such voids:

$$\frac{\Delta V}{V_0} = \beta |P_E| = \frac{4}{3}\pi R^3 n \qquad (3)$$

The typical radius of growing cavitation nanovoids $R$ in the vicinity of the tip electrode during a voltage pulse does not exceed tens of nanometers [3]. Each nanovoid is periodically polarized in an alternating electromagnetic field and can scatter radiation like particles. Since $R \ll$ $R \ll \lambda/m$ is satisfied for nanovoids of radius $R$, scattering from cavitation nanovoids occurs in the Rayleigh regime, as from small particles with polarizability [15,16]:

$$\alpha_R = 4\pi\varepsilon_0 R^3 \left|\frac{1-m^2}{1+2m^2}\right| \qquad (4)$$

Here λ is the wavelength of the laser used for diagnostics, $m \cong 1.334$ is the real part of the refractive index of water for visible light. For the estimations, we will neglect the dispersion of the sizes of nanovoids that appeared at different times and developed under a variable electrostrictive pressure. The typical size of the region occupied by nanovoids is ~ 1 mm. The radiation scattering cross section $\sigma$ on such discontinuities in the Rayleigh regime can be calculated by the formula [18]

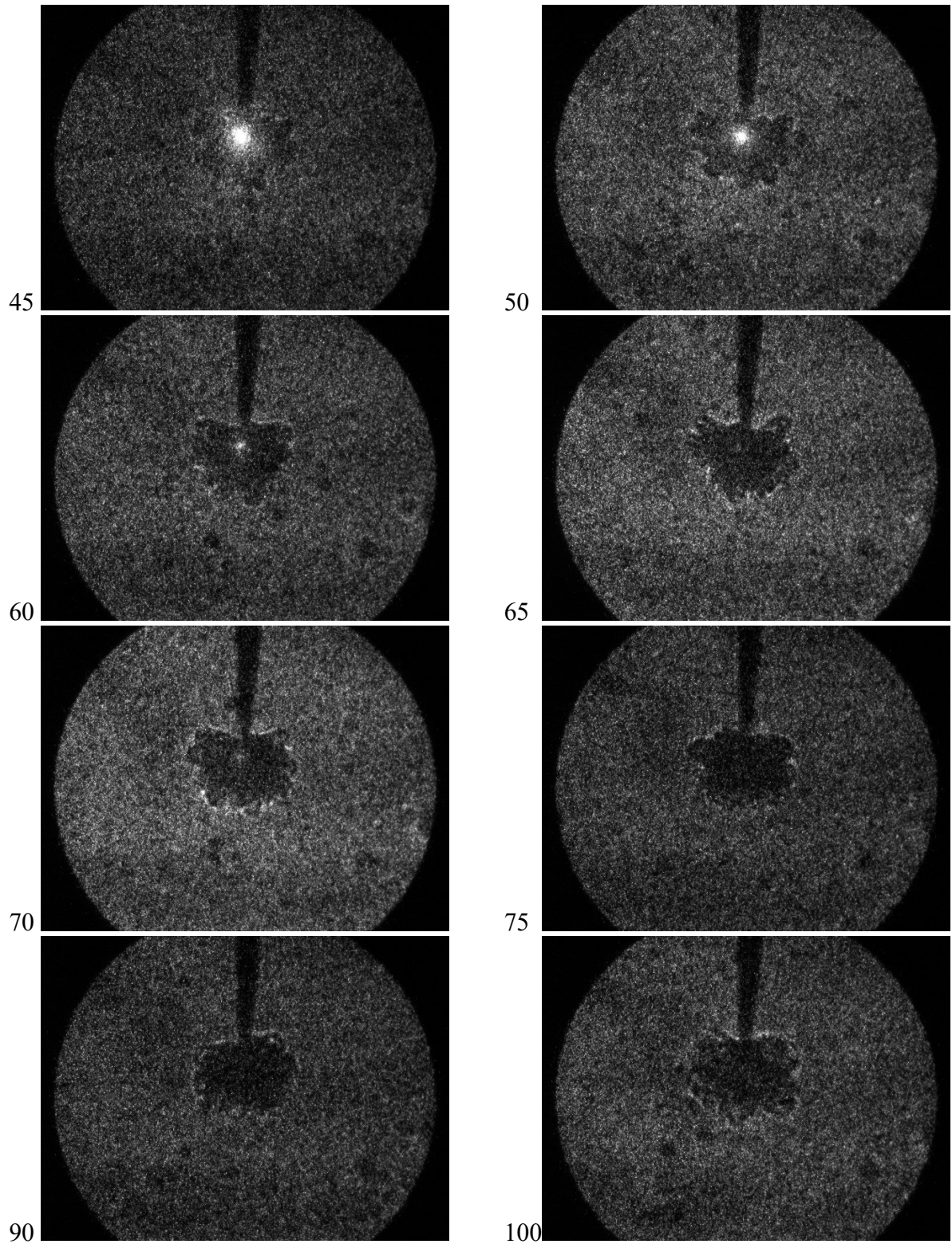

Figure 7. Shadowgraph image of the discharge development zone. The image size is 9.0×6.7 mm$^2$. The camera gate is 2 ns, the illumination is by a laser diode with a wavelength of 520 nm. The numbers indicate the time in nanoseconds from the start of the discharge.

$$\sigma = \frac{8\pi^3 \alpha_R^2}{3\varepsilon_0^2 \lambda^4}. \tag{5}$$

Substituting the polarizability expression (4) into (5), we obtain the Rayleigh scattering cross section of a single spherical nanovoid

$$\sigma = \frac{128\pi^5}{3\lambda^4} R^6 \left|\frac{1-m^2}{1+2m^2}\right|^2. \tag{6}$$

Since, for a laser with a wavelength of $\lambda = 520$ nm, used for the measurements in this work, the absorption in the region of interest is negligible, we can assume that the extinction cross section is determined by Rayleigh scattering from the resulting nanovoids. Thus, assuming that the distribution of nanovoids in the scattering region is uniform, the attenuation of the radiation in the discharge region can be described by the dependence

$$I = I_0 \exp(-n\sigma l). \tag{7}$$

Here, $I_0$ is the radiation intensity in the absence of scattering, $n$ is the concentration of scattering centers, and $l$ is the characteristic size of the scattering region.

To quantify the number and size of emerging nanovoids, we estimate the relative attenuation of radiation in the region of high electric fields.

Combining equations (3), (6), (7), the following estimates for the concentration and characteristic radii of the scattering centers can be easily obtained:

$$n = \frac{24\pi^3}{\ln(I_0/I)} \frac{l}{\lambda^4} \left(\frac{\Delta V}{V_0}\right)^2 \left|\frac{1-m^2}{1+2m^2}\right|^2 \tag{8}$$

and

$$R = \left(\frac{\ln(I_0/I)}{32\pi^4} \frac{\lambda^4}{l} \left(\frac{V_0}{\Delta V}\right) \left|\frac{1+2m^2}{1-m^2}\right|^2\right)^{1/3} \tag{9}$$

Figure 8(a) shows the relative attenuation of the probe beam at 520 nm in the near-electrode region 60 ns after the start of the discharge.

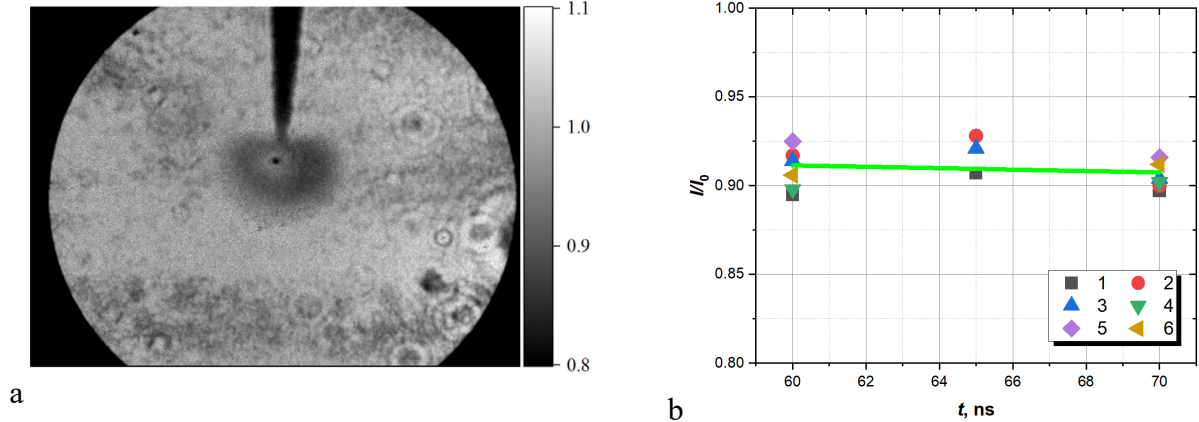

Figure 8. a) Relative attenuation I/I0 of the probe beam at 520 nm in the near-electrode region. The image size is 9.0×6.7 mm² (1392×1080 pixels). The camera gate is 2 ns. Time from the start of the discharge is 60 ns, averaged over 120 pulses. b) Dynamics of the emission attenuation and nanovoids parameters after the nanosecond discharge. 1-3 - initial pulse; 4-6 - secondary pulse. 1,4 - 1st region; 2,5 - 2nd region; 3,6 - 3rd region. Line - linear approximation.

Figure 8(b) shows the dynamics of the radiation scattering intensity variation in the near afterglow of the discharge. To reduce the statistical noise, averaging was performed over an area of 50×50 pixels on the axis of the discharge gap below the tip of the high-voltage electrode. To control for the influence of the choice of averaging area and the number of points included in the processing, Figure 8b also shows the point sets obtained by processing an alternative 50×50 area and an enlarged 100×100 pixel area. The uniformity of the incident intensity distribution $I_0$ was measured in separate experiments (Figure 2). The data shown by points 1-3 correspond to the first high-voltage pulse arriving at the high-voltage electrode through the coaxial cable. The interelectrode distance and voltage were always chosen so that the discharge completely avoided overlapping the gap and the transition to the high-current phase. Due to the high impedance of the discharge gap, the high voltage pulse was almost completely reflected from the discharge cell. As the pulse propagated along the coaxial cable back from the cell, it reached the high voltage generator and was again reflected in the direction of the discharge cell. This secondary pulse reached the high voltage electrode with a delay relative to the first pulse of 70 ns, which is almost exactly the estimate of the doubled signal propagation time in a 2×7 m = 14 m long coaxial cable with polyethylene insulation ($1/c \sim 5$ ns/m). The known delay between pulses allows us to combine the data of the first (points 1-3) and second (points 4-6) pulses in time and compare the data obtained in the afterglow of each pulse. Note that at the interval $t < 60$ ns there is still significant emission from the decaying plasma of the discharge (Fig. 7). For this reason, these points were not included in the consideration. After $t = 60$ ns, the radiation scattering on the nanovoids almost reaches the stationary value $I/I_0 \sim 0.91$ and does not depend on time within the accuracy of the measurements (Figure 8(b)).

The establishment of a quasi-stationary level of radiation scattering at this time interval is determined by the hydrodynamics of the nanovoids. Note that their formation occurs in the region of maximum electric field gradients. The extent of such a region is very small, and this region advances along the gap with the discharge propagation velocity $V \sim 0.1$ mm/ns, (Figure 3), from the tip of the high-voltage electrode. Immediately after the electric field value decreases, the pressure of the ponderomotive forces disappears and the nanovoids begin to collapse. Since the fluid motion in the process of nanovoids formation and in the process of their collapse is caused by the same forces, both processes occur on the nanosecond time scale (the typical time of displacement of a 10 nm thick liquid layer by the thickness of such a layer under the action of 50 MPa pressure is $\tau \sim 0.1$ ns). Thus, the formed nanovoids can exist only in a small region of peak electric fields. However, the collapse of nanovoids cannot occur in the adiabatic regime due to the development of hydrodynamic instabilities, which cause significant residual hydrodynamic perturbations that have the same spatial scale as the initial nanovoids and create optical perturbations in the medium. The time scale of the disappearance of such secondary perturbations is determined by the viscous losses and the thermal conductivity of the liquid. These processes occur at times much longer than the nanosecond range under study, and therefore the parameters of the nanostructures scattering the probe radiation on the inhomogeneities created by the ponderomotive forces practically do not change the spatial scale on this time scale.

Thus, the perturbations of the optical density of the medium observed at nanosecond times are a consequence of nanovoids arising at sub-nanosecond times and their subsequent mechanical collapse. In this case, the spatial scale of the resulting perturbations does not change to a first approximation, and it is possible to estimate the size of the initial nanovoids by analyzing the pattern of perturbations on the interval of tens of nanoseconds.

Obviously, the initial size of the nanovoids and their concentration determine the volume occupied by them (Equation 3). Equating this volume occupied by nanovoids to the decrease in the volume of the liquid phase under the action of ponderomotive forces (Figure 5) and taking into account the data on the scattering of the probing laser (Figure 8b), one can estimate the total volume occupied by nanovoids in the region of maximum ponderomotive forces. Using relations (8), (9), it is easy to obtain the dependence of the characteristic size of nanovoids and their concentration, taking into account the characteristic size of the scattering region of ~ 1 mm (Figure 8a).

Figure 9 shows the results of calculating the characteristic concentration of nanovoids from the scattering data of the test laser radiation at the wavelength $\lambda = 520$ nm (Equation 8). It is easy to see that all points within the measurement scatter are in the interval $n = 1 \div 2 \times 10^{21}$ m$^{-3}$, or $3 \div 6 \times 10^{-5}$ of the number of water molecules in the volume. In the volume of the whole discharge, the total number of nanovoids is $N = 0.5 \div 1 \times 10^{11}$.

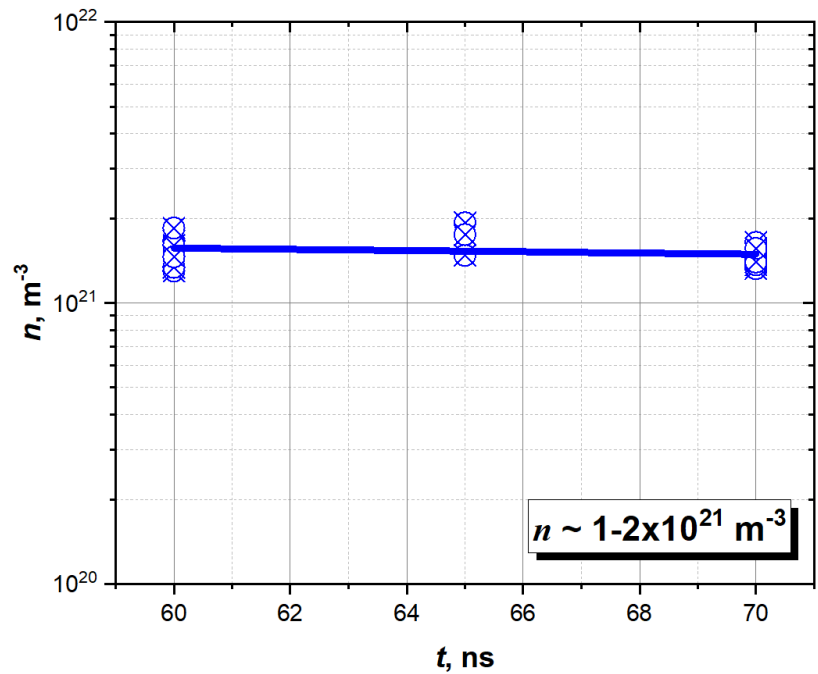

Figure 9. Characteristic concentration of nanovoids in the discharge zone.
The pulse voltage is 30 kV. Points – experiment. Line – linear approximation.

Figure 10 shows the data on the characteristic sizes of the nanovoids resulting from the ponderomotive forces. The data on Rayleigh scattering of radiation at a wavelength of 520 nm with the help of relation (9) allow us to estimate the characteristic radius of the emerging inhomogeneities. It should be noted that this relation assumes sphericity of the nanovoids. At the same time, the ponderomotive forces arising at the ionization wavefront are directed strictly perpendicular to the wavefront and most likely create disc-shaped structures rather than spherical ones, and the arrangement of such nano-disks is strongly correlated. However, there is currently no data on the possible shape of such structures and the peculiarities of Rayleigh light scattering under these conditions. For this reason, the data on the sizes of nanovoids obtained in this work should be considered as an estimate that allows us to analyze threshold processes in the formation of cavitation bubbles.

In view of the above, the calculation using formula (9) allows us to obtain an estimate of the radius of optical density perturbations occurring in the medium as $R \sim 15$ nm. Note that, as in the case of the concentration of nanovoids (Figure 9), the dependence of the perturbation size on the time after discharge is absent within the scatter of experimental data (Figure 10). As mentioned above, exactly such a behavior of nanovoids can be expected in the nanosecond time range, when the rate of dissipative processes (viscosity, thermal conductivity) does not have time to significantly blur the perturbations appearing at sub-nanosecond scale. However, the characteristic size of the nanovoids significantly exceeds the size of water molecules (~0.3 nm) and suggests that the nanovoid formed under the action of ponderomotive forces occupies 50-100 molecular layers. This size allows, with a certain degree of approximation, to apply a macroscopic description to the surface of such a void and to attribute to it, for example, such properties as surface tension. On the other hand, the appreciable size of the void allows it to qualitatively change the picture of the development of the breakdown in the liquid during the development of the pulse discharge. In the case of void formation, an electron emitted from one of its walls travels a collision-free path that is two orders of magnitude longer than its path length in a continuous medium. Thus, there appears to be a mechanism of energy gain by electrons without significant enhancement of the electric field. Based on the obtained estimate of the nanovoid size and the minimum energy required for an electron to ionize a water molecule (in the gas phase, this energy is 12.62 eV; in the liquid phase, the threshold ionization energy is reduced to 10 eV due to interaction with surrounding molecules [20]). In this case, the critical value of the electric field for the development of the discharge in a liquid with a high dielectric permittivity (and therefore a strong effect of ponderomotive forces and nanovoids formed with their help) can be estimated as $E_{crit} \sim \frac{10[V]}{30[nm]} = 3.3 \times 10^8$ V/m, which is in good agreement with the estimate obtained earlier from measurements of the amplitude of the applied voltage and the radius of curvature of the ionization wave propagating in the liquid (equation (2)).

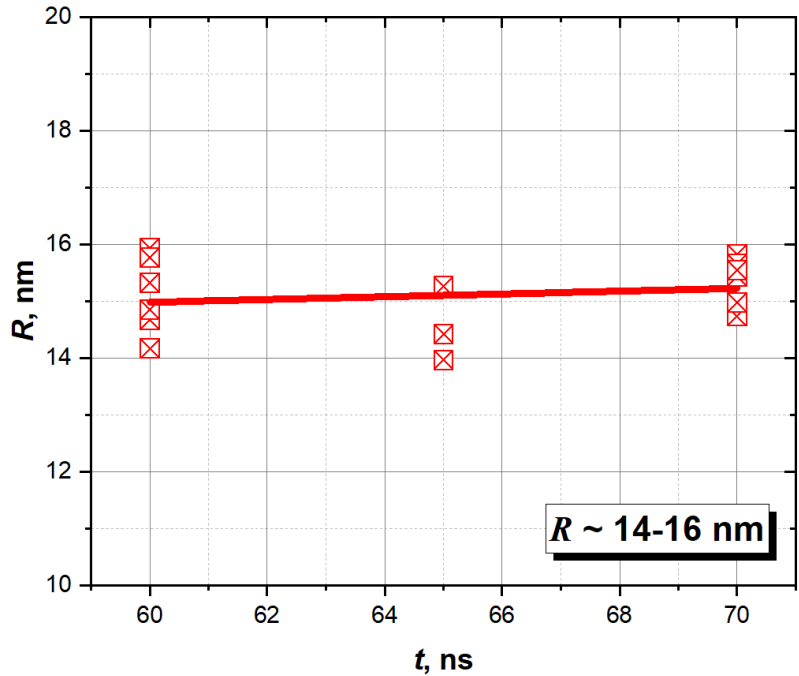

Figure 10. Characteristic radius of nanovoids in the discharge zone.
The pulse voltage is 30 kV. Points - experiment. Line - linear approximation.

Thus, strong electric fields at the ionization front create a relatively high concentration of nanovoids at this point. This effect makes it possible to study the initial stages of cavitation development on the nanosecond time scale and in the nanometer size range. The results obtained are in good agreement with theoretical estimates for similar conditions [16]. It is important to note that in our analysis of the measurement results we have assumed that all cavitation nanovoids are of the same size. However, since the formation and growth of cavitation nanovoids takes place all the time in the region where the negative pressure amplitude exceeds the critical one, a spectrum of cavitation nanovoids of different sizes is formed, depending on the time of their formation [19].

**Conclusions**
The results of the work demonstrated the possibility of using Rayleigh scattering to study and evaluate the parameters of the earliest stages of cavitation development in areas of negative pressure.

The theoretical picture of the development of cavitation and the formation of many nanovoids over sub-nanosecond times in the region of a strong inhomogeneous electric field is confirmed by the experiments conducted in this work.

The formation of nanovoids contributes to the formation of a nanosecond breakdown in a liquid. Under nanosecond electric pulse conditions, the electrostatic forces support the expansion of nanoscale voids behind the front of the ionization wave; in the wave front the extreme electric field provides a strong negative pressure in the dielectric fluid due to the presence of electrostriction forces, forming the initial nanovoids in the continuous medium. An estimate of a critical electric field for breakdown development with nanovoids formation is estimated to be $E_{crit} \sim 3 \times 10^8$ V/m, which is approximately one order of magnitude smaller than the field required for the development of a discharge in a continuous medium [21]. Data on the concentration and size of the nanovoids formed were obtained. It was shown that the strong electric field with a pulse amplitude of 30 kV in this region generates about $N \sim 0.5\text{-}1.0 \times 10^{11}$ nanovoids with a typical number density $n \sim 1.5 \times 10^{21}$ m$^{-3}$ and a characteristic size $R \sim 15$ nm.

**Acknowledgements**

The work was partially support by NSF Award No. 2120400/2129409 "Probing Cavitation Inception in Dielectric Liquids with Sub-Nanosecond Precision".